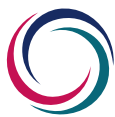

# In situ transport characterization of magnetic states in Nb/Co superconductor/ferromagnet heterostructures

Olena M. Kapran[1], Roman Morari[2,3], Taras Golod[1], Evgenii A. Borodianskyi[1], Vladimir Boian[2], Andrei Prepelita[2], Nikolay Klenov[4,5], Anatoli S. Sidorenko[2,3,6] and Vladimir M. Krasnov[*1,3]

Address:
[1]Department of Physics, Stockholm University, AlbaNova University Center, SE-10691 Stockholm, Sweden, [2]Institute of Electronic Engineering and Nanotechnologies, MD2028 Chisinau, Moldova, [3]Moscow Institute of Physics and Technology, State University, 141700 Dolgoprudny, Russia, [4]Lomonosov Moscow State University, Faculty of Physics, Moscow, 119991, Russia, [5]Moscow Technical University of Communication and Informatics, 111024 Moscow, Russia and [6]Laboratory of Functional Nanostructures, Orel State University named after I.S. Turgenev, 302026, Russia

Email:
Vladimir M. Krasnov[*] - vladimir.krasnov@fysik.su.se

* Corresponding author

## Abstract

Employment of the non-trivial proximity effect in superconductor/ferromagnet (S/F) heterostructures for the creation of novel superconducting devices requires accurate control of magnetic states in complex thin-film multilayers. In this work, we study experimentally in-plane transport properties of microstructured Nb/Co multilayers. We apply various transport characterization techniques, including magnetoresistance, Hall effect, and the first-order-reversal-curves (FORC) analysis. We demonstrate how FORC can be used for detailed in situ characterization of magnetic states. It reveals that upon reduction of the external field, the magnetization in ferromagnetic layers first rotates in a coherent scissor-like manner, then switches abruptly into the antiparallel state and after that splits into the polydomain state, which gradually turns into the opposite parallel state. The polydomain state is manifested by a profound enhancement of resistance caused by a flux-flow phenomenon, triggered by domain stray fields. The scissor state represents the noncollinear magnetic state in which the unconventional odd-frequency spin-triplet order parameter should appear. The non-hysteretic nature of this state allows for reversible tuning of the magnetic orientation. Thus, we identify the range of parameters and the procedure for in situ control of devices based on S/F heterostructures.

## Introduction

Competition between spin-polarized ferromagnetism and spin-singlet superconductivity leads to a variety of interesting phenomena including the possible generation of the odd-frequency spin-triplet order parameter [1-3]. In recent years, this exotic state has been extensively studied both theoretically [4-17] and experimentally [18-35] in various superconductor/ferromagnet (S/F) heterostructures. It is anticipated, that this phenomenon can be employed for the creation of novel superconducting devices, in which the supercurrent is determined and controlled by the magnetic state of the heterostructure, that is, by the relative orientation of magnetizations in several F-layers [18,19,22,23,25-35].

However, the practical realization of such devices is complicated because neither ways of controlling many degrees of freedom in S/F multilayers, nor methods for monitoring magnetic states in micro- or nanoscale S/F devices are established. The situation is complicated by a variety of coexisting phenomena: (i) Both singlet and triplet currents with short- and long-range components can flow through S/F heterostructures [12]. Therefore, even a long-range supercurrent cannot be automatically ascribed to the triplet order. (ii) The supercurrent strongly depends on a usually unknown domain structure in F [23,30,36] and flux quantization in S [37,38], both influenced by size and geometry. (iii) The long-range spin-triplet supercurrent appears only in the noncollinear magnetic state [5,9-12]. Therefore, utilization of this phenomenon for device applications requires accurate determination and control of the micromagnetic state of micro- or nanoscale devices. Similar control is needed for the operations of a large number of superconducting spintronics devices, including memory elements and spin valves [22,29,34,39-43]. The need for establishing experimental characterization techniques for the in situ monitoring of magnetic states in S/F micro- and nanoscale devices is our main motivation.

Here we study experimentally in-plane transport properties of microstructured Nb/Co multilayers (MLs) with a different number of F-layers and layer thicknesses. Our goal is to demonstrate how conventional experimental techniques can be used for in situ assessment of magnetic states of small S/F devices. The key technique that we employ is the first-order-reversal-curves (FORC) analysis. We demonstrate that in combination with magnetoresistance (MR) and Hall effect measurements, it can provide a detailed knowledge of the magnetic configuration in the ML. In particular, we identify the parallel (P) state, the antiparallel (AP) state, the noncollinear monodomain scissor state, and polydomain states. We observe that the domain state is manifested by a profound enhancement of resistance. Analysis of the Hall effect reveals that those maxima are associated with the flux-flow phenomenon, caused by motion of Abrikosov vortices induced by domain stray fields. From a device application perspective, the most important is the noncollinear scissor state, in which the unconventional odd-frequency spin-triplet order parameter should appear. The non-hysteretic nature of this state allows for reversible tuning of the magnetic configuration. Thus, we identify the range of parameters and the procedure for the controllable operation of devices based on S/F heterostructures.

## Samples

We study two types of Nb/Co MLs with different numbers of F-layers and layer thicknesses. The simplest S1, Nb(50 nm)/Co(1.5 nm)/Nb(8 nm)/Co(2.5 nm)/Nb(8 nm)/Si ML (bottom-to-top), has just two dissimilar Co layers composing a single pseudo spin valve. A more complex S2, Nb(50 nm)/[Co(1.5 nm)/Nb(6 nm)/Co(2.5 nm)/Nb(6 nm)]$_3$Co(1.5 nm)/Nb(6 nm)/Si (the structure in square brackets is repeated three times) has five Co layers. MLs are deposited by magnetron sputtering in a single deposition cycle without breaking the vacuum. We use a Nb target (99.95% purity) for deposition of S-layers, Co (99.95% purity) for F-layers, and Si (99.999%) for seeding bottom and protective top layers. MLs are grown on a Si(111) wafer. Prior to deposition, targets were precleaned by plasma-etching for 3 min and in addition for 1 min upon switching between targets. The deposition is performed at room temperature with a water-cooled sample stage. Thicknesses are defined using calibrated growth rates: 3.5 nm/s for Nb and 0.1 nm/s for Co. For every set of F-layers, three identical samples were prepared simultaneously, and some were used for the calibration of the etching rates of the films. MLs are patterned into micrometer-scale bridges with multiple contacts using photolithography and reactive ion etching. A scanning electron microscopy (SEM) image of one of the studied samples is shown in Figure 1a.

Control of the magnetic state implies a possibility of variation of a relative magnetization orientation in neighbor F-layers, which requires different coercive fields. To facilitate this, we use dissimilar Co layers with thicknesses of 1.5 and 2.5 nm. Nb/Co MLs with similar Co thicknesses have been studied earlier and demonstrated good uniformity and perspectives for device applications [25,26,28,32,33,43,44]. Figure 1b shows a magnetization curve for a similar Nb(25 nm)/[Co(1.5 nm)/Nb(8 nm)/Co(2.5 nm)/Nb(8 nm)]$_6$Co(1.5 nm)/Nb(25 nm) unpatterned ML film, deposited using the same setup (data from [43]). $M(H)$ is obtained by SQUID magnetometry in a field parallel to the film in the normal state at $T > T_c$. A significant hysteresis of $M(H)$ reveals the in-plane anisotropy of Co films (albeit with a small coercive field, $H_C \approx 30$ Oe), consistent with earlier studies [44-47].

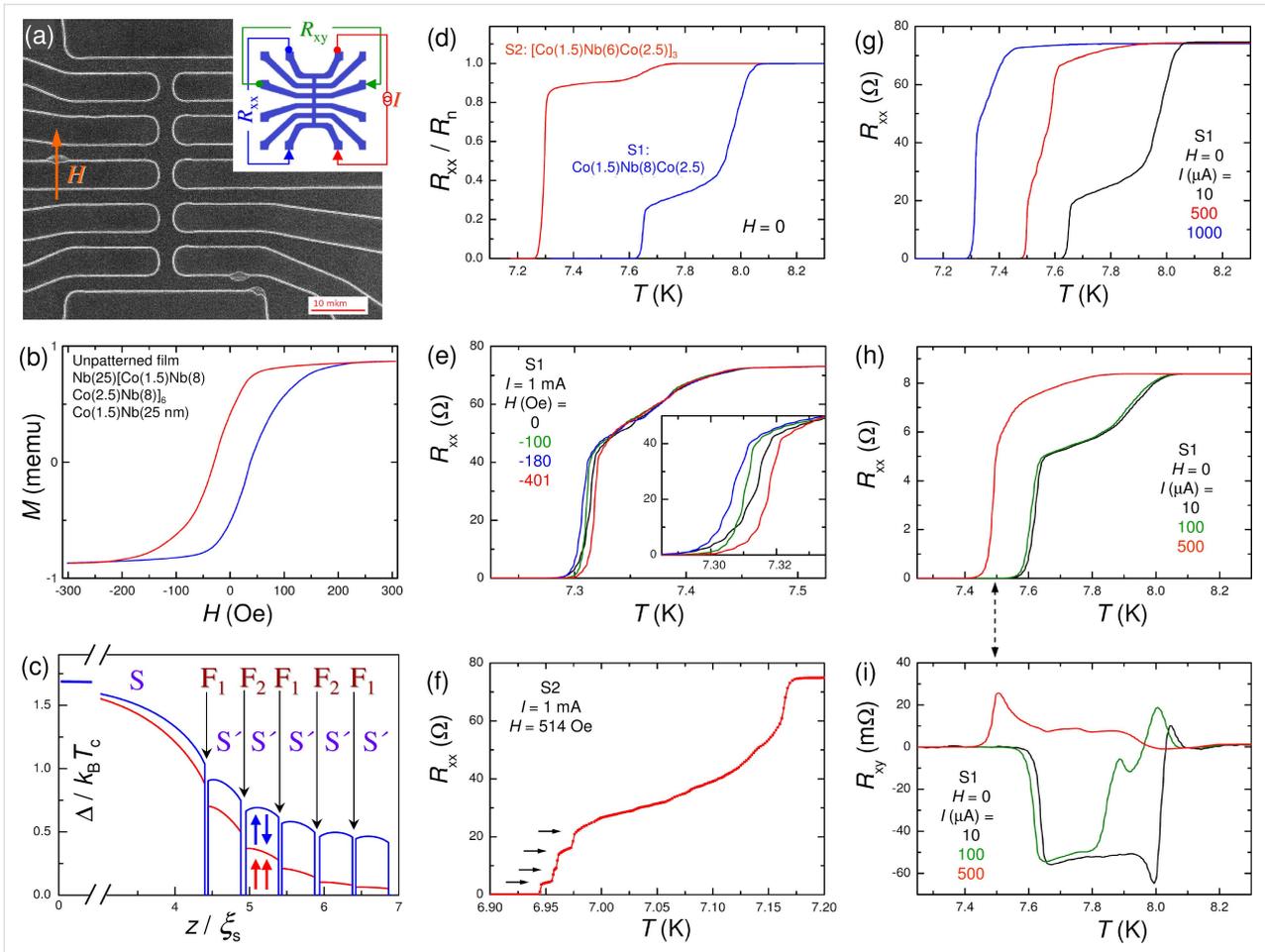

**Figure 1:** (a) SEM image of a micropatterned Nb/Co multilayer. The sample contains twelve contacts, six horizontal and one vertical bridge with widths of a few micrometers. The inset shows the contact configuration for simultaneous measurements of longitudinal and Hall resistances in panels (h) and (i). The orange arrow indicates the orientation of the magnetic field in all experiments. (b) Magnetization curve of an unpatterned Nb(25)[Co(1.5)Nb(8)Co(2.5)Nb(8)]$_6$Co(1.5 nm)Nb(25) ML at $T > T_c$. The blue/red curves represent up/down field sweeps (data from [43]). (c) Simulated superconducting energy gap, $\Delta$, in the S2 ML for P (red) and AP (blue) states. (d) Longitudinal resistance as a function of the temperature, normalized by $R_n$ ($T \geq T_c$), for both MLs. Measurements are taken with small currents at $H = 0$. Double transitions are attributed to different critical temperatures of thick bottom S-layer and thin S′-spacers. (e) $R_{xx}(T)$ for a horizontal bridge (S1) at four consecutively increasing field strengths and $I_{ac} = 1$ mA. The inset demonstrates a nonmonotonous dependence on the magnetic field. (f) $R_{xx}(T)$ for a horizontal bridge (S2) at $H = 514$ Oe and $I_{ac} = 1$ mA. The resistive transition is significantly broadened, compared to the case in (d). Several small steps (marked by arrows) with similar resistance increments are seen. (g) Bias dependence of $R_{xx}(T)$ transitions at $H = 0$ for a horizontal bridge at S1. (h) Longitudinal and (i) Hall resistances for a vertical bridge at the S1 sample at different bias currents and $H = 0$. Contact configuration is shown in the inset in panel (a). A significant and sign-reversal Hall signal is observed within the transition region. This is a fingerprint of the flux-flow phenomenon, caused by the motion of Abrikosov vortices in S-layers.

Figure 1c shows a numerical simulation of the superconducting order parameter, $\Delta$, distribution in a S/F ML similar to S2 [48]. It provides a qualitative understanding of the modulation of the proximity effect in the ML in P (red) and AP (blue) states. A thick bottom S-layer, Nb(50 nm), acts as a Cooper pair reservoir and is only modestly affected by F-layer orientation. However, thin S′ spacers, Nb(6 and 8 nm), with a thickness comparable to the superconducting coherence length, $\xi_S \approx 10$ nm, are strongly affected. Superconductivity in S′ layers is stronger in the AP state and is almost quenched in the outmost S′ layer in the P state. This is caused by subtractive (in the AP state) and additive (in the P state) influences of ferromagnetic exchange fields from neighbor F-layers, which are detrimental to the spin-singlet order in S′ [1-8]. Simulations in Figure 1c demonstrate the tunability of superconductivity in such S/F MLs by changing the magnetic state. Because of the bottom S-layer, there is a gradient of $\Delta$ in S′ layers, which implies that S′ layers have dissimilar superconducting properties.

Measurements are performed in a closed-cycle $^3$He cryostat with a superconducting magnet. More details about experimental setup and sample fabrication and characterization can be found in [23,34] and [43], respectively. Resistances are measured by the lock-in technique with different current amplitudes $I_{ac}$. In all cases, the magnetic field is applied parallel to the film plane in the orientation sketched in Figure 1a. The multiterminal geometry of samples allows for simultaneous four-probe measurements of different segments of the sample in both longitudinal, $R_{xx}$, and Hall, $R_{xy}$, directions. When current is sent through the central vertical bridge, as sketched in Figure 1a, the measurements correspond to the easy-axis magnetization orientation (field along the long side of the vertical line). Alternatively, we can send current through horizontal bridges, which corresponds to the hard-axis magnetization orientation (field perpendicular to the long side of the bridge). In the latter case, the resistance of the whole electrode is measured in a four-probe manner using two separate wires bonded to each contact pad.

## Results

Figure 1d shows $R_{xx}(T)$ dependencies, normalized by the normal state resistance $R_n$ ($T \geq T_c$), for microbridges at S1 (blue) and S2 (red) MLs at $H = 0$. Resistances are measured with $I_{ac} = 10$ μA for S1 and 20 μA for S2, which correspond to approximately equal small current densities in both MLs. Both MLs show a double transition, which could be attributed to different critical temperatures in S and S' layers, $T_c(S') < T_c(S)$. Consistent with this assumption, $T_c$ of the S2 ML with thinner S' (6 nm) is smaller than that of the S1 ML with S' (8 nm).

Figure 1e shows $R_{xx}(T)$ curves for a horizontal bridge at the S1 sample at four sequentially increasing magnetic field strengths (hard axis orientation) and $I_{ac} = 1$ mA. It is seen that the onset of resistivity at $T \approx 7.3$ K is affected by the field. However, this effect is nonmonotonous with the field, as can be seen from the zoomed-in view in the inset. The rest of the transition is little affected in this field range (up to 500 Oe). This is caused by the large value of the upper critical field for thin Nb films in the parallel field [49]. Therefore, the observed nonmonotonous field dependence is not directly induced by the applied field but reflects the remagnetization process of F-layers in the ML.

Figure 1f shows the $R_{xx}(T)$ curve for a horizontal bridge at the S2 sample (hard axis orientation), measured at $H = 514$ Oe with $I_{ac} = 1$ mA. Here, several steps (marked by arrows) with similar resistance increments of ca. 5 Ω can be distinguished close to the onset of $R_{xx}(T)$. They are probably due to individual transitions of the five S' layers in this ML, which have a gradient of the order parameter, as seen from Figure 1c.

### Non-linear flux-flow Hall effect

Resistivity in type-II superconductors with sizes larger than the London penetration depth, λ, is caused by motion of Abrikosov vortices, that is, it has a flux-flow (FF) nature [50-53]. Since our micrometer-size bridges are significantly larger than λ ≈ 100 nm of Nb, the FF phenomenon is anticipated. FF depends on mag-netic field, transport current, and vortex pinning, which in turn depends on the temperature. Figure 1g shows the bias-dependent variation of resistive transitions at $H = 0$ for a horizontal bridge at the S1 sample.

For a single S layer at $H = 0$ vortices may only be induced by the self-field of a transport current. However, for S/F heterostructures they may also be induced by stray fields from F-layers, especially in the presence of domain walls [54-58]. Domain walls also affect the FF phenomenon because they create a pinning landscape for vortices: vortices are pinned to domains and cannot move across them, but can freely move along domain walls [55]. Therefore, we expect that the magnetic state of a ML may influence the FF resistance.

Figure 1h shows $R_{xx}(T)$ for a vertical bridge at the S1 sample measured with the contact configuration depicted in the inset of Figure 1a. All the curves are obtained at $H = 0$, without remagnetizing the ML. However, measurements are made with different current amplitudes $I_{ac} = 10$ μA (black), 100 μA (olive), and 500 μA (red). At low and intermediate bias currents, 10 and 100 μA, a two-step $R_{xx}(T)$ transition occurs with $T_c(S) = 8$ K and $T_c(S') = 7.8$ K and the shape of $R_{xx}(T)$ remains the same. This indicates that the current–voltage characteristics at low bias are almost linear and resistance is almost bias-independent. However, at high bias, $I_{ac} = 500$ μA, $R_{xx}(T)$ is significantly smeared out and the onset shifts to significantly lower $T$. This manifests entrance into the non-linear regime.

Figure 1i shows corresponding Hall resistances measured as shown in the sketch from Figure 1a. The measured Hall signal contains a small (<0.1%) admixture of the longitudinal resistance, which has been subtracted as described in [59]. It is seen that a noticeable Hall signal appears only within the resistive transition region and is non-linear (bias-dependent) even at low bias. Furthermore, at low and intermediate bias $R_{xy}$ changes sign. Such a behavior is typical for the FF Hall effect in superconductors [50-52]. This provides clear evidence for the FF phenomenon in our samples. From the comparison of Figure 1h and Figure 1i it is clear that FF takes place in the whole resistive transition region.

## Flux-flow magnetoresistance, triggered by domains

Figure 2a–c shows longitudinal magnetoresistances at different temperatures. Measurements are done on a horizontal bridge (hard axis orientation) at the S2 sample, the same as in Figure 1f, and at the same current $I_{ac} = 1$ mA. Blue/red lines represent up/down field sweeps. The curves in Figure 2a are measured at the very onset of the resistive transition, as seen by the small (milliohms) resistance values. Here we observe the

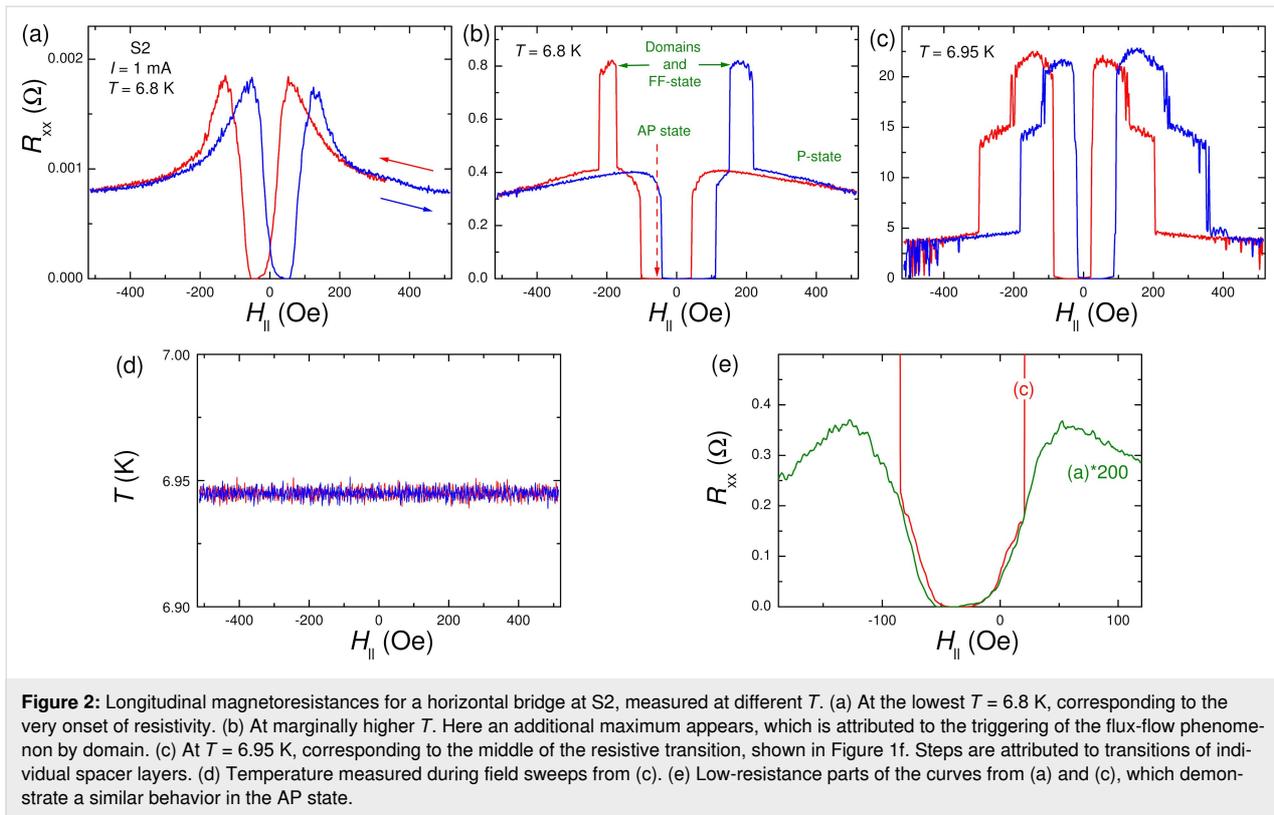

**Figure 2:** Longitudinal magnetoresistances for a horizontal bridge at S2, measured at different *T*. (a) At the lowest *T* = 6.8 K, corresponding to the very onset of resistivity. (b) At marginally higher *T*. Here an additional maximum appears, which is attributed to the triggering of the flux-flow phenomenon by domain. (c) At *T* = 6.95 K, corresponding to the middle of the resistive transition, shown in Figure 1f. Steps are attributed to transitions of individual spacer layers. (d) Temperature measured during field sweeps from (c). (e) Low-resistance parts of the curves from (a) and (c), which demonstrate a similar behavior in the AP state.

simplest MR curves, most consistent with the two-state theoretical prediction based on mono-domain simulations, as in Figure 1c. Namely, a P state at high field with large $R_{xx}$ (suppressed superconductivity) and an AP state at low fields with small $R_{xx}$ (enhanced superconductivity) [4-6,8-10,20,32,34,35]. However, this simple scenario does not explain the appearance of additional maxima between the P and AP states. The discrepancy becomes more pronounced at higher *T*, that is, higher up at the $R_{xx}(T)$ transition.

Figure 2b shows MR at almost the same *T* as in Figure 2a with a higher, but still quite small $R_{xx} < 1\ \Omega$. Figure 2c represents MR at higher *T*, corresponding to the middle of the first transition in $R_{xx}(T)$ from Figure 1f. Here the maxima between P state (high fields) and AP state ($R_{xx} \approx 0$ at low fields) become profound. The appearance of such a maximum also follows from the nonmonotonous field variation of the $R_{xx}(T)$ curves, as can be seen from the inset in Figure 1e. In Figure 2c it is seen that the resistance switches stepwise between certain values of ca. 0, 5, 10 (missing in (c), but accessible at slightly different *T*), 15, and 20 Ω. They correspond to steps in $R_{xx}(T)$ indicated in Figure 1f, which we attributed to transitions of the individual S′ layers. From Figure 2b and Figure 2c it can be seen that the fields at which resistance drops slightly depend on *T*. This does not allow for a straightforward association of such a drop with a transition into the AP state, which should be temperature-inde-pendent in this temperature range. The discussed jumps in MR are not caused by thermal instabilities. This is illustrated in Figure 2d, which represents the sample temperature during field sweeps from Figure 2c. Therefore, variation of MR reflects the transformation of magnetic states in the multilayer.

In Figure 2e we replot low-field parts of the downward curves from Figure 2a and Figure 2c. It is seen that they coincide after proper scaling, that is, they represent a temperature-independent part of MR, which could be associated with the magnetic state. In this case, the minima with $R_{xx}$ near 0 from −20 to −50 Oe should represent the range of existence of the AP state. Observation of this scaling points out that the underlying magnetic state does not depend on *T* and that the difference in shapes of the MR curves is caused by something else.

First of all, as we mentioned in the Introduction, the triplet order parameter can be generated in the noncollinear states between P and AP states [4-6,34,37]. It can lead to a reverse proximity effect due to enhanced leakage of Cooper pairs into F-layers [6,20,21]. This could lead to a weakening of superconductivity in S′ layers and, thus, may explain the appearance of the third high-resistance state between P and AP states [6]. However, there should be yet another explanation for the reported strong temperature dependence of MR, seen in Figure 2a–c, because the remagnetization process of F-layers should be practically independent on temperature in this very narrow temperature range, which is also supported by the scaling, shown in Figure 2e.

The second reason for the appearance of the distinct intermediate state between P and AP states appears if remagnetization is associated with splitting into a polydomain state. The latter is expected for not very small structures [37]. Our micrometer-scale F films are large enough to anticipate such polydomain remagnetization [59]. Since observation of the non-linear Hall effect, Figure 1i, reveals the FF origin of resistivity in our bridges, the observed unusual maximum in MR is likely caused by triggering of the FF phenomenon by domain stray fields, which change upon remagnetization of F-layers. The remarkable temperature variation of MR in Figure 2a–c is then primarily caused by the temperature variation of the vortex mobility. The latter may indeed show strong temperature variations close to $T_c$.

## In situ characterization of magnetic states by the first-order-reversal-curves analysis

The aim of this work is to establish techniques for in situ transport characterization of magnetic states in micropatterned S/F MLs. The key technique that we employ for this purpose is FORC, a powerful tool for the characterization of magnetic states in complex ferromagnetic structures [60-62]. FORC analysis starts at the same saturated state. Then the field is swept to a reversal field $H_r$ and measurements are carried out on the way back to the saturated state. The experiment is repeated with gradually varying $H_r$. Recently, it has been shown that FORC can be used for the analysis of magnetic states in S/F MLs [34]. Essentially, we have to search for the appearance of different types of hysteresis in the FORC response, which indicates switching into some specific metastable magnetic states. There are two mechanisms for the appearance of hysteresis in the MLs [34,37]. The major hysteresis is associated with switching in and out from the magnetostatically stable AP state. Multiple smaller ones are associated with switching between different domain states, which are also metastable.

Figure 3a shows longitudinal MR for the same horizontal bridge (hard axis orientation) at the S1 sample as in Figure 1e. The overall shape of $R_{xx}(H)$ with a minimum at low fields and an additional maximum nearby is similar to that for S2 in Figure 2b. Figure 3b represents the FORC analysis for this bridge. The magnetic field is swept from above the saturation field $H \approx 500$ Oe down to the reversal field $H_r$ and back to the saturation field. Red lines represent forward curves and blue the FORCs. For $H_r \approx 14$ Oe, curve-1 in Figure 3b, the FORC is reversible. The non-hysteretic behavior corresponds to a coherent monodomain rotation of magnetization in a scissor-like manner

[34,37]. At $H_r \approx -11$ Oe, curve-2, the FORC starts to show a tiny signature of hysteresis, which disappears at $H > 50$ Oe. For $H_r \approx -40$ Oe, curve-3, the forward (red) curve reaches the minimum ($R_{xx} \approx 0$) characteristic for the AP state, and the reversal curve (blue) starts to exhibit a clear hysteresis. With further increase of $H_r$ within the minimum, the reversal curve is practically unchanged, as seen from curve-4. This indicates that the state of the ML remains the same. As reported in [34], the appearance of the initial and very stable hysteresis at small fields is associated with switching to the AP state. This is fully consistent with our observation that the initial hysteresis corresponds to the minimum of $R_{xx}$.

With further increase of $H_r$, beyond the AP minimum, the reversal curve clearly changes, see curve-5, indicating switching into a different magnetic state. Furthermore, the range of hysteresis expands to $H > 200$ Oe and several additional small switches occur within this range. The curve-6 in Figure 3b is obtained for $H_r = -160$ Oe in the middle of the $R_{xx}$ maximum. The reversal curve is clearly different from curve-5, revealing yet another initial state. With a further increase of $H_r$ towards the negative saturation field, the hysteresis changes gradually until reaching the saturated state, shown in Figure 3a. Such a gradual transformation indicates that the ML is in a polydomain state with many close metastable states.

For a more quantitative analysis of FORC data, following [34], in Figure 3c we plot the $H_r$-dependence of the normalized hysteresis area between forward and reverse curves, that is, the integral of the absolute value of the difference between the red and blue curves in Figure 3b. Magenta symbols represent the FORC data. They are plotted on a reverse scale along the vertical axis to resemble magnetization curves. From this plot it is clearly seen how the hysteresis starts to develop at $H_r < -10$ Oe and saturates at $H_r < -350$ Oe. However, this curve does not represent the magnetization curve. In particular, at the AP minimum in the range $-120$ Oe $< H_r < -20$ Oe the state of the ML remains the same, as follows from the similarity of FORC curves 3 and 4 in Figure 3b. In this range the hysteresis area is growing linearly with $H_r$ simply because the integration range is increasing as ($H^*$–$H_r$), where $H^*$ is the field for the onset of hysteresis. In Figure 3d we plot hysteresis area divided by ($H^*$–$H_r$). With such a normalization, the AP state is properly described by a plateau. For comparison we also show a sketch of the expected magnetization curve (olive line, right axis), with the AP plateau at $0.25 = (2.5 \text{ nm} - 1.5 \text{ nm})/(2.5 \text{ nm} + 1.5 \text{ nm})$ of the saturation magnetization $M_s$. It is seen that this representation provides a remarkably close description of $M(H)$ curves at intermediate fields. At higher field values, $H < -220$ Oe, the points start to fall down, which indicates the approach to the saturation state at which the hysteresis area does no longer depend on the integration range ($H^*–H_r$), as seen from Figure 3c.

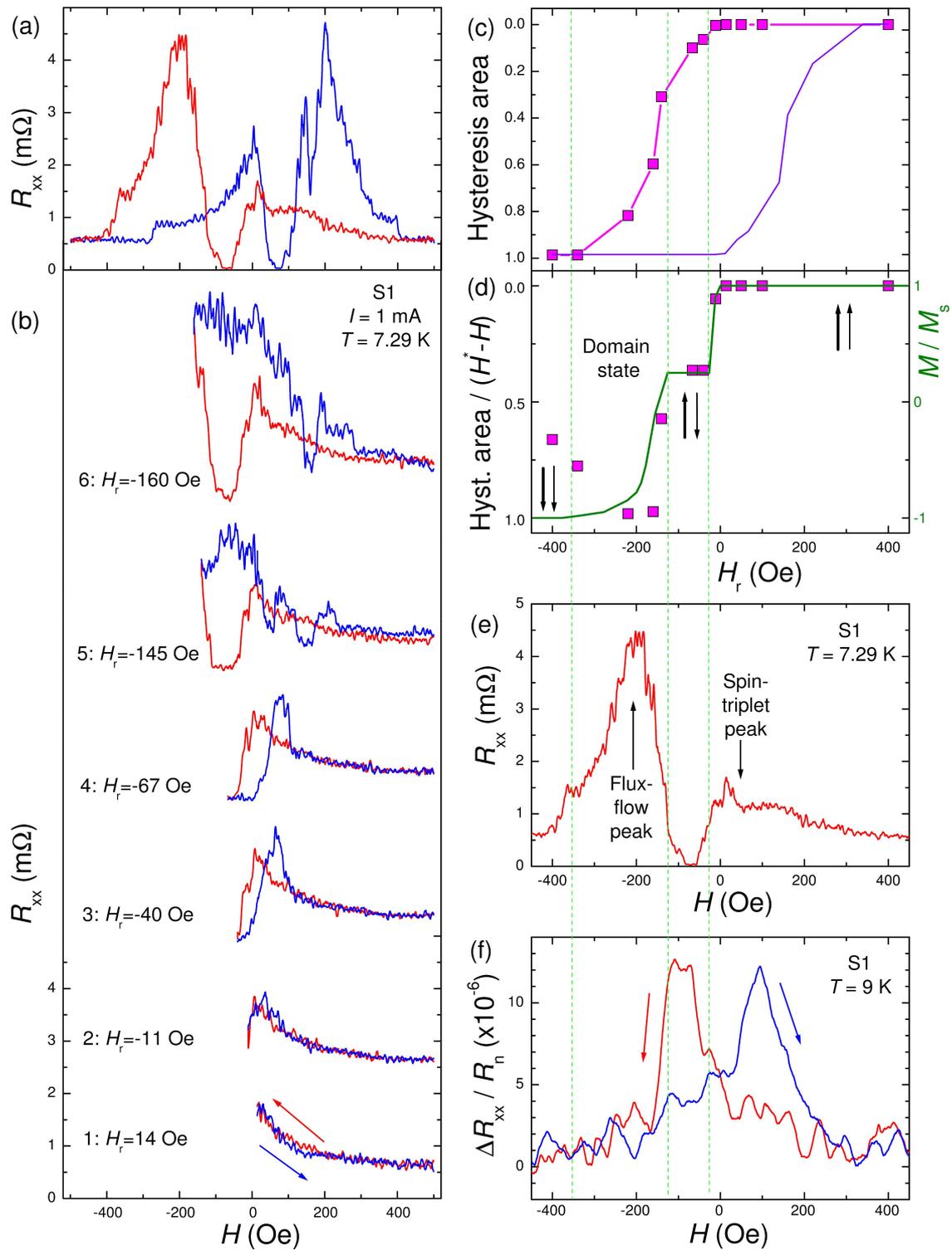

**Figure 3:** (a) Magnetoresistance of a horizontal bridge on the S1 sample, *T* = 7.29 K. (b) FORC analysis of MR for different reversal fields *H*$_r$ (curves are shifted vertically for clarity). (c) Hysteresis area between forward (red) and reversal (blue) curves in (b). The violet curve shows a centrally symmetric reflection of the same data. (d) Hysteresis area normalized by the integration field range *H**-*H*$_r$, where *H** is the field for the onset of the hysteresis. The olive line (right axis) represents a sketch of the expected magnetization curve. (e) Full-scale *R*$_{xx}$(*H*) curve, *T* = 7.29, K for a downward field sweep. Here we attribute the small peak at positive fields to a spin-triplet-proximity effect in the monodomain noncollinear state and the large peak at negative fields to the appearance of the flux-flow phenomenon, triggered by domain stray fields. (f) Normal state MR, (*R*$_{xx}$(*H*) − *R*$_n$)/*R*$_n$, for the same bridge at *T* = 9 K. Green vertical lines emphasize correlations between features in (c–f).

For completeness of the analysis, in Figure 3e, we replot the full MR curve and, in Figure 3f, we show the normal-state MR for the same bridge at $T = 9$ K. The two rightmost vertical lines in Figure 3c–f mark the onset and the end of the AP state. The leftmost vertical lines mark the onset of the saturation. The two rightmost lines emphasize a clear correlation between the onset of hysteresis (Figure 3c,d), the minimum of resistance in the superconducting state (Figure 3e), and the maximum of resistance in the normal state (Figure 3f), which all are signatures of the AP state. The two leftmost vertical lines indicate correlations between the maximum in $R_{xx}$ (Figure 3e) and a gradual transition state between the AP state and the negative P state (Figure 3d,f).

Thus, from FORC analysis we conclude that upon remagnetization of the ML from the P state, it first enters into a coherently rotating, nonhysteretic, scissor state, after which it abruptly switches into the AP state, stays in it for a while, and then breaks into a polydomain state, which triggers the flux-flow phenomenon. With further increase of the field, the polydomain state gradually turns into the opposite P state. This picture is consistent with the assessment based on the Hall effect, Figure 1i, and MR analysis, Figure 2, and with earlier FORC analysis of Ni-based $SF_1NF_2S$ spin valves [34].

## Discussion

As we have seen, information about the evolution of magnetic states in S/F MLs is encoded in the shapes of MR curves. The most remarkable feature of those is the three-state remagnetization with an unusual maximum in addition to well-understood P and AP states. As we already mentioned, the third state with enhanced resistance can be caused both by generation of the spin-triplet order parameter in the noncollinear state via the reverse proximity effect [6,21,22] or by a more trivial effect of domain stray fields upon polydomain remagnetization. However, since the magnetic state of F-layers should not exhibit strong temperature dependence in a very narrow temperature range, in which large variation of MR are observed, see Figure 2a–c, spin-triplet mechanism alone cannot explain our observation, and the domain-triggered FF mechanism should play an important role. Indeed, Hall effect measurements, Figure 1i, unambiguously show that the resistance in our films has a flux-flow origin. Therefore, MR is due to the modulation of FF. The latter depends on the vortex density, pinning, the superconducting order parameter, and the driving current.

Next, we argue that: (i) Vortices in thin films have a pancake structure (Pearl vortices) with a field perpendicular to the film.

Therefore, vortex density depends on the magnetic induction $B_z$ perpendicular to the film. Since the applied field in our experiment is parallel to the film, it does not contribute to $B_z$. (ii) A parallel magnetic field does suppress superconductivity. However, it should cause a monotonous (parabolic) increase of $R_{xx}$ with increasing field, which is not the case. Thus, field variation as such does not explain the observed nonmonotonous MR. (iii) Since all measurements in Figure 2 are done with the same $I_{ac} = 1$ mA, the current self-field is not changing and, therefore, cannot cause modulation of MR. (iv) Remagnetization of F-layers can contribute to the flux-flow phenomenon only if it generates the perpendicular field component. Since our Co layers have in-plane anisotropy [43-47], this is only possible via stray fields from domain walls [54-58]. This brings us to the following interpretation of the observed unusual MR, as indicated in Figure 2b: The MR modulation has two contributions. The first is the conventional suppression/enhancement of superconductivity (order parameter modulation) in the P/AP states at high/small fields [4-6,8-10,20,32,34,35]. The second contribution is caused by the FF phenomenon triggered by domains. Domains both create vortices and generate a pinning landscape that determines the direction of vortex motion [55,58]. Domains could explain the observed dramatic variation of MR shapes with $T$. At low $T$, Figure 2a, the FF is negligible (vortices are almost immobile) and the MR is dominated by order parameter modulation, leading to appearance of a minimum at low fields. With increasing $T$, vortices get depinned and FF modulated by domain texture becomes a dominant factor, causing the appearance of additional maxima between the P and AP states as in Figure 2b and Figure 2c.

Generally, the polydomain state is unwanted in most devices based on S/F heterostructures because it is hard to control. Domains cause an irreversible behavior of the heterostructure, associated with both the magnetic hysteresis and with a generation of Abrikosov vortices, which are pinned at film defects. As shown previously [23], those two factors may dramatically distort the characteristics of S/F devices.

FORC is the key technique that we are advertising for in situ transport characterization of magnetic states. As shown in section "In situ characterization of magnetic states by the first-order-reversal-curves analysis" and in [34], FORC may provide detailed information about the variation of magnetic states in micro- and nanoscale S/F devices. In [34] this was demonstrated using out-of-plane measurements. Here we show that similar information can be obtained using in-plain transport measurements. From FORC analysis in Figure 3 we can characterize the evolution of micromagnetic states in our structures upon remagnetization. We observe that the initial stage of remagnetization from P to AP state is fully reversible, see

curve-1 in Figure 3b. Micromagnetic simulations show that such a stage corresponds to monodomain coherent rotation of magnetizations in neighbor F-layers in opposite directions in a scissor-like manner [34,37]. As seen from Figure 3d, the hysteresis appears abruptly upon switching into the AP state. The AP state is magnetostatically stable and therefore persists in a certain field range. However, switching from AP to the opposite P state occurs gradually, which according to micromagnetic simulations indicates splitting into a polydomain state [34,37]. The more domains, the more gradual is the transition to the saturated state. The appearance of domains triggers the flux-flow state by introducing Abrikosov vortices, as reported before [55-58], and leads to the appearance of the additional FF maximum in MR.

Thus, the domain-triggered FF state strongly masks the coexisting odd-frequency spin-triplet state. The latter, however, can be seen in the reversible part of remagnetization curve, from the P to the AP state, which corresponds to positive fields in Figure 3e. Indeed, reversibility of this state indicates the absence of vortices that always cause irreversibility due to pinning. Therefore, we attribute the smaller peak in MR at positive fields in Figure 3e solely to the spin-triplet mechanism. The profound asymmetry of the peaks at positive and negative fields indicates the presence of an additional FF mechanism at negative fields, corresponding to an irreversible domain-vortex state.

## Conclusion
To conclude, we have studied in-plane transport properties of microstructured Nb/Co multilayers. We demonstrated how conventional "transport" techniques can be used for the assessment of magnetic states of small S/F heterostructures and devices. For this, we apply various experimental techniques, including magnetoresistance, Hall effect, and first-order-reversal-curves analysis. We have shown that a combination of those techniques, performed simultaneously, can provide detailed knowledge about the evolution of micromagnetic states. FORC is the key technique that we advertise for such in situ characterization. Using it we identified the parallel state, the antiparallel state, the monodomain scissor state, and polydomain states. Polydomain states are manifested by a profound enhancement of resistance caused by the flux-flow phenomenon, triggered by domain stray fields.

Importantly, the scissor state corresponds to the noncollinear magnetic state of the multilayer in which the unconventional odd-frequency spin-triplet order parameter should appear in the heterostructure. The nonhysteretic nature of this state allows for the controllable tuning of magnetic orientation. Thus, we identify the range of parameters and the procedure for the controllable operation of superconducting spintronic devices based on

S/F heterostructures. Essentially we conclude that for moderately small (micrometer-scale) devices controllable and highly reversible operation can be achieved at fields between one of the P states down to the AP state without entering into the realm of the opposite P state.

## Acknowledgements
We are grateful to Sergey Bakurskiy, Andrey Schegolev, Yury Khaydukov, Mikhail Kupriyanov, and Alexander Golubov for stimulating discussions.

## Funding
The work was partially supported by the European Union H2020-WIDESPREAD-05-2017-Twinning project "SPIN-TECH" under grant agreement Nr. 810144 (samples preparation), the Russian Science Foundation grant No. 19-19-00594 (V.M.K.: data analysis and manuscript preparation), and by the Russian Science Foundation Grant Nr. 20-62-47009 (R.M., A.S.: experiments at low temperatures and magnetization measurements). The manuscript was written during a sabbatical semester of V.M.K. at MIPT, supported by the Faculty of Science at SU.

## ORCID® iDs
Olena M. Kapran - https://orcid.org/0000-0001-7433-4140
Vladimir Boian - https://orcid.org/0000-0001-5708-4098
Nikolay Klenov - https://orcid.org/0000-0001-6265-3670
Vladimir M. Krasnov - https://orcid.org/0000-0002-3131-8658

## Preprint
A non-peer-reviewed version of this article has been previously published as a preprint: https://arxiv.org/abs/2010.03454

## References
1. Buzdin, A. I.; Vedyayev, A. V.; Ryzhanova, N. V. *Europhys. Lett.* **1999,** *48,* 686–691. doi:10.1209/epl/i1999-00539-0
2. Kadigrobov, A.; Shekhter, R. I.; Jonson, M. *Europhys. Lett.* **2001,** *54,* 394–400. doi:10.1209/epl/i2001-00107-2
3. Bergeret, F. S.; Volkov, A. F.; Efetov, K. B. *Phys. Rev. Lett.* **2001,** *86,* 3140–3143. doi:10.1103/physrevlett.86.3140
4. Buzdin, A. I. *Rev. Mod. Phys.* **2005,** *77,* 935–976. doi:10.1103/revmodphys.77.935
5. Bergeret, F. S.; Volkov, A. F.; Efetov, K. B. *Rev. Mod. Phys.* **2005,** *77,* 1321–1373. doi:10.1103/revmodphys.77.1321
6. Fominov, Y. V.; Golubov, A. A.; Karminskaya, T. Y.; Kupriyanov, M. Y.; Deminov, R. G.; Tagirov, L. R. *JETP Lett.* **2010,** *91,* 308–313. doi:10.1134/s002136401006010x
7. Blanter, Y. M.; Hekking, F. W. J. *Phys. Rev. B* **2004,** *69,* 024525. doi:10.1103/physrevb.69.024525
8. Eschrig, M. *Rep. Prog. Phys.* **2015,** *78,* 104501. doi:10.1088/0034-4885/78/10/104501
9. Houzet, M.; Buzdin, A. I. *Phys. Rev. B* **2007,** *76,* 060504(R). doi:10.1103/physrevb.76.060504

10. Asano, Y.; Sawa, Y.; Tanaka, Y.; Golubov, A. A. *Phys. Rev. B* **2007,** *76,* 224525. doi:10.1103/physrevb.76.224525

11. Trifunovic, L.; Popović, Z.; Radović, Z. *Phys. Rev. B* **2011,** *84,* 064511. doi:10.1103/physrevb.84.064511

12. Mel'nikov, A. S.; Samokhvalov, A. V.; Kuznetsova, S. M.; Buzdin, A. I. *Phys. Rev. Lett.* **2012,** *109,* 237006. doi:10.1103/physrevlett.109.237006

13. Pugach, N. G.; Buzdin, A. I. *Appl. Phys. Lett.* **2012,** *101,* 242602. doi:10.1063/1.4769900

14. Alidoust, M.; Sewell, G.; Linder, J. *Phys. Rev. Lett.* **2012,** *108,* 037001. doi:10.1103/physrevlett.108.037001

15. Richard, C.; Buzdin, A.; Houzet, M.; Meyer, J. S. *Phys. Rev. B* **2015,** *92,* 094509. doi:10.1103/physrevb.92.094509

16. Hikino, S.; Yunoki, S. *Phys. Rev. B* **2015,** *92,* 024512. doi:10.1103/physrevb.92.024512

17. Meng, H.; Wu, J.; Wu, X.; Ren, M.; Ren, Y. *Sci. Rep.* **2016,** *6,* 21308. doi:10.1038/srep21308

18. Bell, C.; Burnell, G.; Leung, C. W.; Tarte, E. J.; Kang, D.-J.; Blamire, M. G. *Appl. Phys. Lett.* **2004,** *84,* 1153–1155. doi:10.1063/1.1646217

19. Robinson, J. W. A.; Halász, G. B.; Buzdin, A. I.; Blamire, M. G. *Phys. Rev. Lett.* **2010,** *104,* 207001. doi:10.1103/physrevlett.104.207001

20. Leksin, P. V.; Garif'yanov, N. N.; Garifullin, I. A.; Schumann, J.; Kataev, V.; Schmidt, O. G.; Büchner, B. *Phys. Rev. Lett.* **2011,** *106,* 067005. doi:10.1103/physrevlett.106.067005

21. Zdravkov, V. I.; Kehrle, J.; Obermeier, G.; Lenk, D.; Krug von Nidda, H.-A.; Müller, C.; Kupriyanov, M. Y.; Sidorenko, A. S.; Horn, S.; Tidecks, R.; Tagirov, L. R. *Phys. Rev. B* **2013,** *87,* 144507. doi:10.1103/physrevb.87.144507

22. Baek, B.; Rippard, W. H.; Benz, S. P.; Russek, S. E.; Dresselhaus, P. D. *Nat. Commun.* **2014,** *5,* 3888. doi:10.1038/ncomms4888

23. Iovan, A.; Golod, T.; Krasnov, V. M. *Phys. Rev. B* **2014,** *90,* 134514. doi:10.1103/physrevb.90.134514

24. Khaydukov, Y. N.; Ovsyannikov, G. A.; Sheyerman, A. E.; Constantinian, K. Y.; Mustafa, L.; Keller, T.; Uribe-Laverde, M. A.; Kislinskii, Y. V.; Shadrin, A. V.; Kalaboukhov, A.; Keimer, B.; Winkler, D. *Phys. Rev. B* **2014,** *90,* 035130. doi:10.1103/physrevb.90.035130

25. Robinson, J. W. A.; Witt, J. D. S.; Blamire, M. G. *Science* **2010,** *329,* 59–61. doi:10.1126/science.1189246

26. Khaire, T. S.; Khasawneh, M. A.; Pratt, W. P., Jr.; Birge, N. O. *Phys. Rev. Lett.* **2010,** *104,* 137002. doi:10.1103/physrevlett.104.137002

27. Banerjee, N.; Robinson, J. W. A.; Blamire, M. G. *Nat. Commun.* **2014,** *5,* 4771. doi:10.1038/ncomms5771

28. Martinez, W. M.; Pratt, W. P., Jr.; Birge, N. O. *Phys. Rev. Lett.* **2016,** *116,* 077001. doi:10.1103/physrevlett.116.077001

29. Glick, J. A.; Aguilar, V.; Gougam, A. B.; Niedzielski, B. M.; Gingrich, E. C.; Loloee, R.; Pratt, W. P., Jr.; Birge, N. O. *Sci. Adv.* **2018,** *4,* eaat9457. doi:10.1126/sciadv.aat9457

30. Lahabi, K.; Amundsen, M.; Ouassou, J. A.; Beukers, E.; Pleijster, M.; Linder, J.; Alkemade, P.; Aarts, J. *Nat. Commun.* **2017,** *8,* 2056. doi:10.1038/s41467-017-02236-2

31. Vávra, O.; Soni, R.; Petraru, A.; Himmel, N.; Vávra, I.; Fabian, J.; Kohlstedt, H.; Strunk, C. *AIP Adv.* **2017,** *7,* 025008. doi:10.1063/1.4976822

32. Lenk, D.; Zdravkov, V. I.; Kehrle, J.-M.; Obermeier, G.; Ullrich, A.; Morari, R.; Krug von Nidda, H.-A.; Müller, C.; Kupriyanov, M. Y.; Sidorenko, A. S.; Horn, S.; Deminov, R. G.; Tagirov, L. R.; Tidecks, R. *Beilstein J. Nanotechnol.* **2016,** *7,* 957–969. doi:10.3762/bjnano.7.88

33. Lenk, D.; Morari, R.; Zdravkov, V. I.; Ullrich, A.; Khaydukov, Y.; Obermeier, G.; Müller, C.; Sidorenko, A. S.; Krug von Nidda, H.-A.; Horn, S.; Tagirov, L. R.; Tidecks, R. *Phys. Rev. B* **2017,** *96,* 184521. doi:10.1103/physrevb.96.184521

34. Kapran, O. M.; Iovan, A.; Golod, T.; Krasnov, V. M. *Phys. Rev. Res.* **2020,** *2,* 013167. doi:10.1103/physrevresearch.2.013167

35. Cascales, J. P.; Takamura, Y.; Stephen, G. M.; Heiman, D.; Bergeret, F. S.; Moodera, J. S. *Appl. Phys. Lett.* **2019,** *114,* 022601. doi:10.1063/1.5050382

36. Weides, M. *Appl. Phys. Lett.* **2008,** *93,* 052502. doi:10.1063/1.2967873

37. Iovan, A.; Krasnov, V. M. *Phys. Rev. B* **2017,** *96,* 014511. doi:10.1103/physrevb.96.014511

38. Golovchanskiy, I. A.; Bol'ginov, V. V.; Stolyarov, V. S.; Abramov, N. N.; Ben Hamida, A.; Emelyanova, O. V.; Stolyarov, B. S.; Kupriyanov, M. Y.; Golubov, A. A.; Ryazanov, V. V. *Phys. Rev. B* **2016,** *94,* 214514. doi:10.1103/physrevb.94.214514

39. Abd El Qader, M.; Singh, R. K.; Galvin, S. N.; Yu, L.; Rowell, J. M.; Newman, N. *Appl. Phys. Lett.* **2014,** *104,* 022602. doi:10.1063/1.4862195

40. Bakurskiy, S.; Klenov, N. V.; Soloviev, I. I.; Kupriyanov, M. Y.; Golubov, A. A. *Appl. Phys. Lett.* **2016,** *108,* 042602. doi:10.1063/1.4940440

41. Bakurskiy, S.; Klenov, N. V.; Soloviev, I. I.; Pugach, N. G.; Kupriyanov, M. Y.; Golubov, A. A. *Appl. Phys. Lett.* **2018,** *113,* 082602. doi:10.1063/1.5045490

42. Shafraniuk, S. E.; Nevirkovets, I. P.; Mukhanov, O. A. *Phys. Rev. Appl.* **2019,** *11,* 064018. doi:10.1103/physrevapplied.11.064018

43. Klenov, N.; Khaydukov, Y.; Bakurskiy, S.; Morari, R.; Soloviev, I.; Boian, V.; Keller, T.; Kupriyanov, M.; Sidorenko, A.; Keimer, B. *Beilstein J. Nanotechnol.* **2019,** *10,* 833–839. doi:10.3762/bjnano.10.83

44. Jeon, K.-R.; Montiel, X.; Komori, S.; Ciccarelli, C.; Haigh, J.; Kurebayashi, H.; Cohen, L. F.; Chan, A. K.; Stenning, K. D.; Lee, C.-M.; Blamire, M. G.; Robinson, J. W. A. *Phys. Rev. X* **2020,** *10,* 031020. doi:10.1103/physrevx.10.031020

45. Parkin, S. S. P.; More, N.; Roche, K. P. *Phys. Rev. Lett.* **1990,** *64,* 2304–2307. doi:10.1103/physrevlett.64.2304

46. Mosca, D. H.; Petroff, F.; Fert, A.; Schroeder, P. A.; Pratt, W. P., Jr.; Laloee, R. *J. Magn. Magn. Mater.* **1991,** *94,* L1–L5. doi:10.1016/0304-8853(91)90102-g

47. den Broeder, F. J. A.; Hoving, W.; Bloemen, P. J. H. *J. Magn. Magn. Mater.* **1991,** *93,* 562–570. doi:10.1016/0304-8853(91)90404-x

48. Bakurskiy, S.; Kupriyanov, M.; Klenov, N. V.; Soloviev, I.; Schegolev, A.; Morari, R.; Khaydukov, Y.; Sidorenko, A. S. *Beilstein J. Nanotechnol.* **2020,** *11,* 1336–1345. doi:10.3762/bjnano.11.118

49. Zeinali, A.; Golod, T.; Krasnov, V. M. *Phys. Rev. B* **2016,** *94,* 214506. doi:10.1103/physrevb.94.214506

50. Samoilov, A. V.; Legris, A.; Rullier-Albenque, F.; Lejay, P.; Bouffard, S.; Ivanov, Z. G.; Johansson, L.-G. *Phys. Rev. Lett.* **1995,** *74,* 2351–2354. doi:10.1103/physrevlett.74.2351

51. Krasnov, V. M.; Logvenov, G. Y. *Phys. C (Amsterdam, Neth.)* **1997,** *274,* 286–294. doi:10.1016/s0921-4534(96)00672-7

52. Sonin, E. B. *Phys. Rev. B* **1997,** *55,* 485–501. doi:10.1103/physrevb.55.485

53. Embon, L.; Anahory, Y.; Jelić, Ž. L.; Lachman, E. O.; Myasoedov, Y.;
    Huber, M. E.; Mikitik, G. P.; Silhanek, A. V.; Milošević, M. V.;
    Gurevich, A.; Zeldov, E. *Nat. Commun.* **2017,** *8,* 85.
    doi:10.1038/s41467-017-00089-3
54. Laiho, R.; Lähderanta, E.; Sonin, E. B.; Traito, K. B. *Phys. Rev. B* **2003,**
    *67,* 144522. doi:10.1103/physrevb.67.144522
55. Vlasko-Vlasov, V. K.; Welp, U.; Imre, A.; Rosenmann, D.; Pearson, J.;
    Kwok, W. K. *Phys. Rev. B* **2008,** *78,* 214511.
    doi:10.1103/physrevb.78.214511
56. Aladyshkin, A. Y.; Silhanek, A. V.; Gillijns, W.; Moshchalkov, V. V.
    *Supercond. Sci. Technol.* **2009,** *22,* 053001.
    doi:10.1088/0953-2048/22/5/053001
57. Iavarone, M.; Scarfato, A.; Bobba, F.; Longobardi, M.; Karapetrov, G.;
    Novosad, V.; Yefremenko, V.; Giubileo, F.; Cucolo, A. M. *Phys. Rev. B*
    **2011,** *84,* 024506. doi:10.1103/physrevb.84.024506
58. Stellhorn, A.; Sarkar, A.; Kentzinger, E.; Barthel, J.; Di Bernardo, A.;
    Nandi, S.; Zakalek, P.; Schubert, J.; Brückel, T. *New J. Phys.* **2020,** *22,*
    093001. doi:10.1088/1367-2630/abaa02
59. Golod, T.; Rydh, A.; Krasnov, V. M. *J. Appl. Phys.* **2011,** *110,* 033909.
    doi:10.1063/1.3615959
60. Béron, F.; Ménard, D.; Yelon, A. *J. Appl. Phys.* **2008,** *103,* 07D908.
    doi:10.1063/1.2830955
61. Dobrotă, C.-I.; Stancu, A. *J. Appl. Phys.* **2013,** *113,* 043928.
    doi:10.1063/1.4789613
62. Dumas, R. K.; Greene, P. K.; Gilbert, D. A.; Ye, L.; Zha, C.;
    Åkerman, J.; Liu, K. *Phys. Rev. B* **2014,** *90,* 104410.
    doi:10.1103/physrevb.90.104410

## License and Terms